\documentstyle[12pt]{article}

\clearpage

\begin{document}
\baselineskip 24pt
\begin{center}
{\Large\bf Superconductivity at Very Hygh  Temperatures - 
Hyperconductivity}\\[5mm]
{\bf V.A.Vdovenkov}
\end{center}
\baselineskip 16pt
\begin{center}
 Moscow State Institute of Radioengineering,
 Electronics and Automation 

(Technical university)\\
 Vernadsky~ave. 78,~117454~Moscow,~Russia\\
\end{center}
\begin{abstract}
The mechanism of superconductivity caused by the electron-vibrational 
centres and their inherent oscillations in crystals and solid-state 
structures near  room temperature  and at higher temperatures - 
hyperconductivity is discussed and actualized. The experimental data about
hiperconductivity in some semiconductors are presented. Hyperconductivity 
arises and exists from normal-superconducting temperature $T_c$ to 800K 
(probably up to temperature of a crystal melting and is higher). 
$T_c$ in various materials  varies from hundred up to several hundreds 
degrees on Kelvin scale depending on size of average nuclear number 
of a material. 
\end{abstract}

The phenomenon of superconductivity was open in 1911 \cite{Onn11}
and till now was observed in various materials when their temperature 
below normal-superconducting transition temperature $T_c$,  density of  
electrical current and magnetic intensity are below 
critical meanings $J_k$ and $H_k$ \cite{Gin77,Gin92}. The meanings 
$J_k$ and $H_k$ tend to zero at increasing of 
temperature and her approaching  to the $T_c$. The first superconductors 
had small meanings of the $T_c$: 4.1K~(Hg), 7.3K~(Pb). To the present time 
many classes of superconductors are open. The greatest meanings of the 
$T_c$ have layered like perovskite metal oxide. $T_c$ has reached 
134K...136K. The critical meanings of the $T_c$, $J_k$ and $H_k$ 
limit applications of superconductors. In this connection the 
increasing of the $T_c$  up to room temperature and above is an 
essential problem. Solution of the problem would allow 
superconductive devices to work at room temperatures 
without cooling. This important problem while was solving 
experimentally, by selection of chemical structure and technological 
processings of materials with the purpose of achievement of high meanings 
of $T_c$. Till now there was no reliance in possibility or 
impossibility of superconductivity near to room temperature or at 
higher temperatures. 

The expectations of superconductivity just at temperatures less than 
$T_c$ are traditionally based on available experimental data and according 
to known physical mechanisms which was called for explanation of 
superconductivity. Among these mechanisms the leading position 
probably occupies the mechanism underlying the known theory of 
superconductivity (BCS)~\cite{Bar57} and based on interaction 
between electrons by means of an interchange of  phonons 
(which nevertheless does not explain high meanings of the $T_c$ 
noticed on experience). According to this theory   
electrons exchanges with each other by virtual phonons that gives rise to  
an attraction between them which  determines magnitude of $T_c$. 
For increasing of the $T_c$ was offered to use  phonons with large energies 
as virtual  phonons (Larkin~A.I.) but unfortunately such mechanism of 
superconductivity was not carried out.

The opportunity of existence in the solid materials of inherent 
oscillations and waves of $\alpha$-~, $\beta$-  and  $\gamma$-  
types is known  \cite{Vdo96,Vdo99}. 
The elementary quantums of inherent oscillations and waves of $\alpha-$ 
type considerably exceed energy of  acoustic and optical phonons. 
The inherent oscillations effectively cooperate with electrons (holes) 
and crystal phonons providing  strong electron-phonon coupling in 
crystals and structures. Such properties of inherent oscillations and 
waves have given hope to use them together with crystal phonons as 
virtual phonons for realization of superconductivity at very high 
temperatures.

The purpose of the given work is to investigate an opportunity of inherent 
oscillations and waves for realization of superconductivity.

The researches were executed on samples of silicon monocrystal  because 
the silicon is rather well investigated, his many characteristics and 
parameters are known, Si for a long time is used as a model material 
and widely is applied in electronics. In particular it is known about 
an opportunity of existence in silicon of the electron-vibrational 
centres (EVC) whose presence is essentially important for realization 
of superconductivity. Also in experiments  were used  samples on the 
basis of other semiconductors: InP, InAs, InSb, Ge, CdHgTe in whose  
EVC supposedly are formed by impurity atoms of oxygen.

\section*{1.~Some properties of inherent oscillations and waves}
\hspace{1.5em}The traditional models of molecules and crystals in 
which the atoms are replaced by physical points with masses of atoms do not 
correspond to adiabatic  Born-Oppenheimer's theory \cite{Born27} 
and do not give opportunities to study inherent oscillations. 
The opportunity of inherent oscillations and waves existence
in the condensed materials can be proved by using of adiabatic  
model of  solids \cite{Vdo99}. 
In such model each atom is submitted as a nucleus and electron 
shell connected with each other  by elastic force and capable 
to carry out oscillatory movements ($\alpha-$ type of oscillations).
Each atom in adiabatic model of a crystal can be submitted as 
inherent  $\alpha-$ oscillator consisting from a nucleus and electron shell.
Elastic force of inherent oscillator is almost proportional to displacement 
of a nucleus from the centre of an electron shell. The value of 
appropriate factor  depend on electrical potential field which is
created by electron shell  near  her centre. The frequency of  
inherent oscillations depends on conditions in which there is 
an atom. In particular the collective properties of electrons 
in a crystal can be expressed so that their displacements 
(displacements of electron shells) are  coherent  (synchronous), 
within  the some length ($\lambda$)  and the coherent  area  acquire 
superconductive properties. The size $\lambda$ can exceed tens of space 
lattice and to achieve significant size. In such conditions 
the nucleus of each atom has an opportunity to oscillate relatively  
common mass of coherent  area. The amplitude  of a nucleus 
oscillations is insignificant $(\sim10^{-2}A^0)$. 

The energy spectrum of 
($\alpha-$type) inherent oscillations in harmonic approach can be presented 
by the expression
\begin{equation}
\label{1} 
E_\upsilon=E_\alpha(1/2+\upsilon)                                             
\end{equation}
where oscillatory quantum number $\upsilon$ = 0, 1, 2..., the  elementary 
oscillatory quantum is defined by mass of a nucleus $M_z$ with 
number Z: $E_\alpha=\hbar(\zeta/M_z)^{1/2}$, $\zeta$- factor connecting 
force with 
displacement of a nucleus from an equilibrim position  
at the centre of an electronic envelop, $\hbar$=h/$(2\pi)$, h - Plank's constant. 
According to accounts (conformable to experimental data) at 
increasing Z magnitude  $E_\alpha$ decreases from 0.402 eV at Z = 2 up to 
0.22 eV at Z = 8 and then increase up to 0.406 eV at Z = 80. 
The inherent oscillations of a $\beta$- type caused by joint displacements 
of  nucleus and K - electron subshell relatively other part of an electron 
shell and also inherent oscillations of a $\gamma$- type caused by joint 
displacement of a nucleus, K -  and L - electron subshells relatively other 
part of an electron shell of atom are possible also.
The elementary quantums of inherent oscillations $\beta$- and 
$\gamma$- types are less than the appropriate quantums of 
$\alpha$- oscillations: $E_\alpha > E_\beta > E_\gamma$. 
Thus  energy of some inherent oscillations with small $\upsilon$ 
in a crystal can not exceed 
energies of the solved electron transitions between various 
electronic states i and j: $E_\upsilon < | E_i-E_j|$. This inequality is a 
condition of applicability an adiabatic principle  
in correspondence with \cite{Dav73}. 
Thus in adiabatic approach in the condensed materials the inherent 
oscillations $\alpha$-, $\beta$- and $\gamma$- types can exist and  
propagate. 
They represents the oscillatory displacement of nucleuses relatively 
electronic environments. The speeds  of inherent oscillations 
waves can achieve speed of a sound.

The theoretical opportunity of existence of  inherent oscillations  
and waves followed from adiabatic model of a crystal, based on 
adiabatic  Born-Oppenheimer's principle, is confirmed by experimental 
data about temperature dependence of crystals resistivity, about 
phonon drag at high temperatures and about optical electron-vibrational 
spectra formed with participation of inherent oscillations. 
The  energy levels of inherent oscillations are shown as deep  energy 
levels and the  electron transitions between these levels or on these 
levels or from these levels are electron-vibrational.  
Electron-vibrational transitions are
accompanied by absorption or emission several (on the average S) 
crystal phonons, where S - constant of electron-phonon interaction.

The following ways of excitation of inherent oscillations can be 
practically important: 

- thermal, at the expense of  phonons energy;

- optical, at the expense of  absorbed photons energy;

- shock , for the account of kinetic  energy of (hot) electrons or holes;

- recombination , at the expense of energy recombination ;

- combined, as a combination of two or several specified ways.

In semiconductors it is convenient to raise inherent oscillations 
at the expense of recombination energy of electrons (holes). For 
this purpose is necessary rather strong interaction of electrons 
with lattice oscillations. The constant S equal to an average number of
phonons participating in the act  of  electron-phonon interaction and 
serves a measure of electron interactions  with phonons. The size 
S in ideal  silicon monocrystal calculated on the data on meaning 
of deformation potential constants  does not exceed 0.03 for 
optical phonons and 0.15 for acoustic phonons \cite{Lax55, Hard62}. 
Hence excitation of inherent oscillations at the expense of interaction 
with electrons in ideal  crystals is rare event  that are in 
agreement with  adiabatic theory according to which electrons 
and the oscillations of a lattice in an ideal crystal do not 
interact with each other.
The effective production of inherent oscillations at the expense 
of recombination energy  is possible in crystals containing such 
defects of a lattice for which the strong interaction of electrons 
with phonons is characteristic. These defects have received the 
name of the electron-vibrational centres (EVC). EVC are the  
Jan-Teller's centres. They represents local defects of a crystal 
which equilibrum position or frequencies of oscillations 
depends on their electronic condition. For EVC the large meanings 
S ($S>1$) are characteristic.  According to the theory 
\cite{Hua50}-\cite{ORo53} the quantity of S 
is close to an average phonons participating in electron-vibrational 
transition to energy  levels of EVC. By the theoretical estimations S can 
exceed 50.
The inherent oscillations of atoms of the basic substance can 
exist and to spread in crystals and crystal structures but 
effectively to raise such fluctuations and waves at the expense 
of electrons or holes energy is possible at presence EVC.
In this connection the inherent oscillations and waves caused by 
atoms of the basic substance and also inherent oscillations and waves 
caused by impurity atoms are possible in crystals.
\section*{2. Impurity  oxygen atoms in silicon}
\hspace{1.5em} Impurity oxygen atoms in  silicon monocrystals are 
electrical inactive and their presence can be established on 
characteristic optical absorption on length of a wave about 
9~micrometer \cite{Kai56}-\cite{Hro57}.The intensity of the 
specified band of absorption reflects the contents of 
oxygen atoms on an optical path in a crystal. The increase of 
concentration of oxygen in silicon is accompanied by occurrence 
of quartz  disseminations  and by displacements of a maximum of the 
characteristic optical band of absorption to 10...11micrometers \cite{Abe66}. 
This  absorption is characteristic for  Si oxides and it is explained 
traditionally in the literature  by optical excitation of chains  
Si-O-Si oscillations. 
In the monography \cite{Vla70} (p. 179) was  shown convincingly that 
the reason of the specified optical band which determine a kind of 
a spectrum  in the field of 9...11 micrometers consist not simply in 
specified chains of atoms but is contained in structure  of a 
lattice (probably in structural defects containing atoms of oxygen). 
The similar bands 
of optical absorption in the specified spectral area are characteristic 
for oxides of InAs, InSb, InP \cite{Vdo81} and oxides of other 
semiconductors that also allows to connect these bands with 
defects of a lattice 
containing atoms of oxygen. Moreover there are weighty bases to 
believe that these defects containing atoms of oxygen are the  
Jan-Teller's centres and they are EVC.

 The presence of EVC inherent oscillators  determine an optical 
properties of crystals.  The oscillatory model describing 
optical spectra of crystals and supposing interaction of 
oscillators with a wide set of crystal phonons is stated 
in \cite{Hua51}. In accordance 
to \cite{Hua51, Born54} the optical reflectivity
\begin{equation}
\label{2}
R_\infty=\frac{(n-1)^2 + k^2}{(n+1)^2 + k^2},
\end{equation}  
where n - optical factor of refraction and k - parameter of absorption  
are defined from the following expressions:
\begin{equation}
\label{3}
n^2-k^2 = \varepsilon_{opt}+\omega^2_p\frac{\Omega^2-\omega^2}{(\Omega^2-\omega^2)^2+\omega^2\nu^2};
\end{equation} 
\begin{equation}
\label{4}
2nk = \omega^2_p \frac{\omega\nu}{(\Omega^2-\omega^2)^2+\omega^2\nu^2}, 
\end{equation}        
where: $\Omega$ - frequency of oscillator, 
$\nu$ - frequency of phonon, 
$\omega$ - optical frequency, $\omega_p=\sqrt{Ne^2/(M\varepsilon_0)}$, 
N - concentration of oscillators, e - charge 
of oscillator, M - mass of oscillator, $\varepsilon_0$ - electrical 
constant, $\varepsilon_{opt}$~-~optical dielectric permeability. 
The optical reflection spectra of silicon oxides and reflection 
spectra of specified semiconductors oxides can be presented as 
sum of separate oscillators reflection spectra. Such decomposition 
of a spectrum corresponds to the quantum theory \cite{Ros51,Nor58}. 
These theories can be applied to the analysis of IR reflection by 
inherent oscillators of EVC.
 
Every separate oscillator have "zero" frequency of oxygen inherent 
$\alpha $ - oscillators ($\hbar\Omega$~=~0.11 eV) which cooperate 
mainly with various lattice 
frequencies between  $\Omega$  and $ \omega _ p $. 
The greatest contribution in  oxydes reflection 
spectra is brought by frequencies appropriate to "zero" 
fluctuations of inherent oxygen oscillators and phonons for 
which ($\Omega /\omega_p$) = 0.25, $\varepsilon_{opt}$=1.2, 
G/$\Omega$=0.011, G - factor of attenuation. The rather 
small meaning $\varepsilon_{opt}$ allows to carry him only to the 
local centre but not to all crystal.
The experimental spectrum of polarized light reflection  
and calculated on the Eq.~(\ref{2}-\ref{4}) 
spectrum of reflection  are given on fig. 1. The experimental spectrum 
was measured at a corner of fall of the linearly polarized radiation 
of 45 degrees for oriented along an axis C quartz monocrystal.
The comparison of spectra submitted in a fig.~1 specifies presence of 
oscillators in quartz with oscillatory energy   which is equal to 
energy of "zero" oscillations of  inherent oxygen oscillators, 
calculated on the Eq.~(\ref{1}) at $\upsilon$ = 0 for atom of oxygen.
Such transitions with frequency of "zero" oscillations are forbidden 
for quantum oscillator but EVC are capable to show duality 
of properties [6]. 
The conditions in a minimum  of oscillator potential and transitions 
with frequency of "zero" oscillations for EVC are accordingly allowable. 
The excess of experimental absorption above calculated absorption at energies 
of quantums more than 0.11~eV is determined by the contribution 
with participation of frequencies $ \omega _ p$ conterminous 
with frequencies of inherent $\alpha$- oxygen oscillators  
calculated on the Eq.~(\ref{1}) at other meanings of~$\upsilon$. 
The same conclusion can be made as a result of similar 
approach of reflection spectra for melted  quartz,  silicate glasses 
and oxides of a number of semiconductors. According to this result 
the optical band which is characteristic for quartz  and others 
oxides of semiconductors is possible to identify with excitation 
of "zero" inherent oscillations of oxygen atom ($\upsilon$ = 0).

Impurity  atoms of oxygen in silicon   irradiated by radiation 
formes associations with vacancies of a crystal lattice known as 
A-centres. As a result of study of  electron spin resonance spectra 
and his dependence from orientation data and about energy levels of 
defects, items of information on a situation of oxygen atoms in   
silicon lattice in \cite{Bem59} the model of A-centre was constructed. 
A-centre represents  anisotropic defect whose electrical dipole 
moment is directed lengthways [110]. A-centre have deep 
electron-vibrational energy levels in the forbidden energy zone 
of silicon which are distant from boundaries ($E_c$ and $E_v$) of the 
forbidden energy zone on sizes multiple  0.11 eV and are close 
to calculated on the Eq. (\ref{1}) for oxygen atom. It allows 
to identify them with inherent $\alpha$-oscillations of oxygen atoms. 
Acceptor  $E_c$~-~(0.16...0.22)~eV and donor $E_v$~+~(0.27...0.33)~eV 
levels are most active at low concentration of A - centres. These 
levels are close to  the calculated energy levels of inherent 
oscillations of oxygen atom. Specified changes  of the energy levels 
are explained by change of complete 
oscillatory energy  of A - centre  caused by the interaction of 
the centres with each other at changing of their concentration 
from $5\cdot10^{17} cm^{-3}$ up to $10^{13} cm^{-3}$ and depends 
on average distance (R) between them as $R^{-3}$. It   coordinates 
with aeolotropic  
structure of A - centre and specifies one-dimensional character 
of interaction between them. The given result corresponds to 
oscillatory model of A-centre as one-dimensional oscillator that 
justifies application Eq.~(\ref{1}) for the description of  his 
electron-vibrational energy levels.

The application of the linearly polarized radiation and  monocrystals 
with rather low concentration of EVC allows to create necessary 
conditions for measurement of electron-vibrational spectra with 
participation of one phonons type. Such spectra may be analysed on 
the basis of the Pekar-Huang- Rhys theory  \cite{Hua50} - \cite{ORo53}.
The typical spectra of photoconductivity and optical transmittance 
connected with photoionization  of  A - centre by polarized IR  
radiation with an electrical vector E $\parallel$ [110] are given on a 
fig. 2. One can see on fig. 2 that spectra are modulated by  phonons. 
The minima of optical transmittance (maxima of optical absorption) 
correspond to minima of photoconductivity. The established conformity 
of extremums in spectra of photoconductivity and absorption \cite{Aki68} 
is explained by characteristic negative photoconductivity when 
electron-vibrational transitions (from valence band on A - centre) 
occur and proves the fact of auto localization of electrons and 
holes on EVC simultaneously. If EVC interacts only with one type 
ocillations of a lattice (similar to  spectra on a fig. 2) then in 
according to \cite{ORo53} the spectral distribution of electron-vibrational 
transitions depends from S,  function Bose  ($f$), contains product  
of the modified  Bessel's function of the order p on 
$\sum_n\delta$(n-p) , where n = 1, 2,... and $\delta$ - delta function, 
p - number of phonons 
participating in electron-vibrational transition. Because of 
presence  $\delta$ - functions p accepts only integer meaning.   p$>$0 
corresponds to absorption of optical  quantum and p$<$0 corresponds 
to radiation of optical  quantum with emission of p phonons. 
Accordingly spectrum contains two wings adequate electron-vibrational  
transitions with absorption and emission of phonons. One wing 
lays below than energy of transition without phonons (p$=$0) 
another wing lays  higher than energy of transition without phonons. 
The wing of a spectrum appropriate to absorption phonons disappears with 
downturn of temperature . At participation of one type 
phonons the spectrum represents a series of discrete lines which 
energy differ on size multiple phonons. Expression for spectral 
distribution includes only one parameter S. Size of S may be selected  
to approach conveniently a contour of an experimental spectrum and 
thus to determine S. The number of phonons which are emitted by 
the centre in a maximum of an optical band is equal S. The energy 
level on which the electron-vibrational transition is carried out 
corresponds to transition without phonons (p = 0).
In the field of high temperatures when $ 4S[f (f + 1)]^{1/2} > p > 1 $ 
the discrete lines extend the periodic structure in spectra disappears 
and the contour of a spectral band  can be approached by function of Gauss:  
\begin{equation}
\label{5}
G(\hbar\omega) = exp{\left\{-\frac{1}{4S[f(f+1)]}(\frac{\omega-\omega_{max}}{\omega_0})^2\right\}}
\end{equation}                                                        
where $\omega$- optical frequency, $\omega_{max}$ - frequency in a 
maximum of 
spectral distribution, $\omega_0$ - optical frequency of electronic 
transition at à = 0.
In the region of low temperatures when f$\longrightarrow$0 
and à $>>$ Sf(f+1) 
contour  of a spectral band follow the dependence 
\begin{equation}
\label{6}
G(\hbar\omega) = S^p/p!.
\end{equation}  
Approach to spectra on basis  of expressions (\ref{5}, \ref{6})  
has  allowed us 
to determine S = 5 for A - centre, types and energy 
of phonons connected with the centres, energy levels of A - centre 
corresponding to transitions with p = 0. In particular, data 
about TA phonons received from the analysis of experimental 
spectra connected with A- centre are given in Table 1 together 
with the appropriate literary data.

~~~~~~

\vspace{0.3cm}
~~~~~~~~~~~~~~~~~\tablename{~1: Energy of TA phonons in silicon}
\vspace{0.3cm}

 ~~~~~~~~~~~~~~~~\begin{tabular}{|c|c|c|c|c|}
\hline
Direction  
&
\multicolumn{4}{|c|}{Energy of phonons (meV) determined by methods:}\\
\cline{2-5}
of phonon & calculating & dissipation & indirect & photocon-\\
wave  & ~~~~ & of nutrons & IR & ductivity\\
vector & \cite{Gor63} & \cite{Mor57} & absorbtion & on A-centre \\
\hline
100 & 23.0 & 21.0 & 22.0 & 22.0 \\
110 & 18.0 & 17.9 & 18.0 & 18.3 \\
111 & 16.5 & 16.7 & 17.0 & 16.9 \\
\hline
\end{tabular}
\vspace{0.3cm}

On an insertion of a fig. 2 are given the experimental data about 
energy splitting of connected with A- centre LO and TA phonons 
arising as a result of interaction A- centres among themselves 
depending on their concentration (N). The size S also changes 
from 5 up to 2 at increase N up to $5\cdot10^{17} cm^{-3}$. Electrons and 
holes carrying out transitions to levels EVC inevitably cooperate   
with p=S phonons. In result at energy levels of EVC, which are 
inherent oscillatory energy levels of these centres and are shown as deep 
energy levels, appears located electrons, holes and phonons. 
Hence EVC represents complex formation consisting from impurity 
atom, his inherent oscillations , electrons, holes and phonons. 
These particles in structure EVC form the interconnected auto  
localized  system submitting  other statistical and dynamic laws  
versus  free phonons and   electrons of conductivity. In this 
connection EVC can cause unusual physical properties to crystals 
and crystal structures which are poorly connected to a structure 
of energy bands of a crystal and properties  of conductivity electrons. 
The similar conclusion was made earlier as a result of research of 
phonon drag  in semiconductors where the phonon drag is caused only by 
those phonons which are strongly connected with electrons and 
consequently have other properties \cite{Fre53, Herr58}.

The modulation of spectra by phonons stops in samples on the basis 
of various semiconductors when the concentration of free charge 
carriers   at room temperature (concentration of doping impurity) 
exceeds $ n _ {max} $ = $2\cdot10 ^ {17} cm ^ {-3}$. It gives the 
basis to consider  that concentration  of electrons  and holes 
which are auto located on EVC and capable to be superconducting 
does not exceed $ n _ {max} $.

A- centre arise at technological processings of silicon and 
silicon structures, in particular, in structures metal - oxide 
of silicon - silicon (MOS-structures). The presence of A- centres 
in structures is shown on characteristic negative photoconductivity 
in silicon under oxide, modulation of spectra by phonons and on 
kinetics of photo-emf  which is described by not less than  triad 
of time constant  (t). These  constants are mutually connected by 
proportions $t_1: t_2: t_3 = 1! /S^1: 2! /S^2: 3! /S^3$ at S = 2 also  
corresponds to probability of electron-vibrational transitions 
which are described by Eq. (\ref{6}) at participation p = 1, 2 and 3 
phonons.
A - centres also influence on  volt - capacitance  characteristics 
of MOS-struktures.
On the fig. 3 are submitted typical volt - capacitance  (C-V)  
characteristics of aluminium contact by the area 
$4.9\cdot10^{-4} cm^2$ to 
a plate of silicon containing A- centres in the concentration
$\sim5\cdot10^{15} cm^{-3}$. The characteristics are measured at room 
temperatureT ($T<T_c$) on various frequencies. The submitted 
characteristics are depended  from frequency. The  dependence 
of the characteristics from frequency is defined by available 
A~-~centres in samples. On various frequencies the  capacitance
is nonmonotonic function of  back  bias that does not correspond 
to traditional theory of capacitance for ideal contact  metal - semiconductor 
with Schottky barrier. Untraditional physical model of contact 
metal - semiconductor is necessary for the analysis of such 
C-V  dependences.
It is possible to approximate the experimental C-V curves 
submitted in a fig. 3 by known dependence of contact   capacitance
(C) from the enclosed voltage (V): 
$C = [S(V)\cdot\varepsilon\varepsilon_0]/L$, 
where $\varepsilon$ - relative dielectric permeability of the semiconductor 
and $e_0$ - electrical constant, L - thickness of depletion region, 
accounting that the effective area of contact (S) depends on a 
voltage: S = S(V). Such dependences corresponds  of contact  
model which is taking into account the presence of fine regions (embedments) 
with high áonductivity in the semiconductor under a field 
electrode which are coherent areas. In process of penetration 
of an electrical field into the semiconductor (at increase of 
back-biasing voltage) the border of depletion region achieves 
some of these embedments and area of equipotential surface  grows. 
Thus differential  capacitance of contact is accordingly increased.
On experience the smooth increase of  capacitance is observed at 
achievement of some bias voltage  that it is possible to explain
by volumetric distribution of the embedments  which serially  
achieves equipotential surface of depletion region border. 
The unmonotonous dependence C(V) with several extremums in 
such model can be explained by presence of several layers of 
embedments. Identical high-resistance and low-resistance layers, 
as it is well known, exists in such and similar structures 
strongly influences conductivity by elastic waves arising in 
depletion region of contact. Besides it is known that the 
periodicity of layers also depends on a voltage on contact 
and from voltage frequency. It brings contribution in frequency 
dependence of contact  capacitance. The appropriate model of contact 
which allow to explain frequency dependence of  capacitance contains 
coherence areas of the small size in the semiconductor under a 
field electrode. These areas are distributed disorderly and are 
grouped in layers parallel to a field electrode. It is possible 
to explain the experimental C-V curves just by a discrete 
structure of layers formated  by fine coherence areas. Monolithic 
low-resistance layers are unsuitable because in this case the 
differential  capacitance can not change in relation to  capacitance of 
depletion region  of the semiconductor. The presence of embedments 
with conductive (superconductive) pieces in the semiconductor under 
field contact quite corresponds to an opportunity of formation 
of coherence areas with the characteristic size $\lambda$. Experimentally 
observable frequency dependence of differential  capacitance in 
semiconductor contacts confirms presence of  coherence areas 
in samples with electron-vibrational centres.

\section*{3. Estimation of parameters }
\hspace{1.5em} Superconductive  property of a material are 
doubtlessly connected to occurrence of coherence areas which possess 
of superconductive properties due to coherence of electronic 
displacements. The law of occurrence of coherence areas  is defined 
by concentration those EVC in which the inherent oscillations 
are exited and also by concentration of phonons. The interaction 
between EVC determines occurrence of coherence areas  and is executed 
by means of an exchanging of the centres with each other by phonons. 
It is possible to assume that for an estimation of the minimal 
concentration ($N_{min}$) of EVC at which the formation of coherence 
areas is possible when such interaction is effective on distances 
between EVC about length of phonon wave  ($\Lambda$). 
The minimal concentration 
of exited EVC can be estimated as $\Lambda^{-3}$. 
Accepting the maximal speed 
of a sound in silicon $V = 9.13\cdot10^5$~cm/c and phonon frequency
$F=1.25\cdot10^{10}~c^{-1}$ we shall 
receive $N_{min} = 2.56\cdot10^{12}~cm^{-3}$.
 
Normal-superconducting transition temperature ($T_c$) can be 
estimated  proceeding from the following reasons. 
If to consider the intrinsic semiconductor material  with effective 
mass for density of conditions for electrons ($m_{nd}$) and 
holes ($m_{pd}$) when at certain temperature ($T$) speed of thermal 
generation of electrons and holes (G) in individual volume and at 
time of life ($\tau_i$) according to the statistical theory define 
inherent concentration 
$n_i$  = $\sqrt{N_cN_v}exp(-\frac{E_g}{2kT})$ = G$\tau_i$. 
$N_c=2(\frac{2\pi m_{nd}kT}{h^2})^{3/2}$ 
and 
$N_v=2(\frac{2\pi m_{nd}kT}{h^2})^{3/2}$  - effective number of 
conditions in conduction 
energy band  and in  valence energy band, $E_g$ - width of energy 
forbidden gape of the semiconductor. In a stationary condition 
for the same time $\tau_i$ carriers of charges recombinate with the same 
rate G. If recombination occurs on energy levels of EVC then 
each recombination act gives rise to EVC inherent oscillations  
with energy determined by Eq.~(\ref{1}) and else  S  phonons is rised. 
Let's consider that it is enough for synchronization of electronic
displacement in coherence areas of one phonon  per one EVC. 
Therefore for formation coherence area  is necessary not less 
$n_ª=(N_{min}/S)\cdot f(E)$ recombination acts  in time $\tau_i$, 
where $f(E)$ - the 
function Bose  for given oscillation energy $E(\mu)$. It is possible 
to write down a condition for existence of the minimal concentration of  
exited EVC as $n_i = n_ª$. By substituting in this equality the 
expressions for $n_i$ and $n_k$we shall receive a relation which is possible to
consider as the equation for definition of temperature $T = T_c$:
\begin{equation}
\label{7}
\sqrt{N_c N_v}exp{(-\frac{E_g}{2kT})} = \frac{N_{min}}{S(exp{\frac{E(\mu)}{kT}}-1)}  ,
\end{equation}
where E($\mu$) - discrete oscillatory energy of EVC. It is necessary 
to tell more about possible meanings of E($\mu$). E($\mu$) can not 
exceed Eg/2  as in this case she will be transferred with a high 
probability to electrons and is spent on electronic transitions 
through the forbidden zone of the semiconductor. It is known  EVC 
show quantum and classical properties [6]: E($\mu$) can accept 
meanings conterminous with energies of charmonical quantum vibrator 
according to Eq.~(\ref{1}) and also meanings multiple $E_\alpha$. 
Therefore for $\alpha$-type inherent EVC oscillator 
E($\mu$) = $(E_\alpha/2)\cdot\mu < Eg/2$, $\mu$ = 1,2,3,...                                         

Using known tabulared meanings of parameters and also S = 5, 
$\upsilon$=0 and $E_\alpha$ = 0.22 í' for A- centre in silicon 
by numerical way is received from (7) the meaning of the $T_c$=305K 
at $\mu$=1, $T_c$=235K at $\mu$=2 and $T_c$=162K at $\mu$=3 for silicon. 
The similar estimation for Ge with the oxygen centres gives meaning 
of the $T_c$=182K, $T_c$=130K and $T_c$=33K. According to this 
results the superconductivity in 
silicon with $\mu=2$ , $E(\mu) = E_\alpha$, for A- centres  necessary 
to expect at temperatures above 235K and in germanium - at 
temperatures is higher 130K.

The estimation of critical  density of current  ($J_k$) can be 
executed proceeding from the data about possible concentration  
of superconducting charges (n) and about maximal their   drift  
speed (v).  The meaning of n can be in 
limits from n = $ n _ k $ up to $n_{max}$ = 2$\cdot 10 ^ {17} cm ^ {-3}$ 
and the speed of drift is probably limited by dissipation on borders of 
coherence areas  and can not exceed speed of a sound V. 
Believing $ J _ k $ = enV we come to the following estimation 
of critical density of a current in silicon at room 
temperature: $0,55 €/cm ^ 2 <  J _ k  < 73 A/cm ^ 2$. 

For an estimation of critical magnetic intensity it is possible 
to involve the following experimental data received at constant density 
of a current in silicon samples in a magnetic field.  Such data are 
given on fig. 4 and represent typical dependence of derivative of  
voltage V (V - on sample)  with respect to magnetic induction (B), 
that is dV/dB from B when current density $ J $ = const, 
measured in a normal 
condition  of Si sample (at $T <  T _ c$) containing EVC with 
concentration $5\cdot10 ^ {15} cm ^ -3$.  The dependence submitted 
on fig. 4 contains periodic maxima and can be explained by 
influence of a magnetic field on charged oscillators as electron 
connected with EVC by means of crystal phonon. The movement 
equation of   such oscillator can be written down   as follows:
\begin{equation}
\label{8}
\frac{d^2X}{dt^2} + (\frac{e}{m^*})\frac{dX}{dt}B - \omega^2 X = 0,
\end{equation}
where X - generalized coordinate , e - charge and m* - effective 
mass of electron, $\omega $ - cyclic frequency of fluctuations, 
B - magnetic induction. Component containing 
$ \frac {dX} {dt} $  takes into account action of Lorentz's 
force  when $ \frac {dX} {dt} $ and B is mutually orthogonal.
The given equation (8) supposes oscillating  solution only under 
condition:
 $\left\{4\omega^2 - [\frac{eB}{m^*}]^2\right\} > 0$.

The oscillatory movements of oscillator are possible when 
$ B < \frac {2m^{*}\omega} {e} $ 
that is when B is not too great. Otherwise oscillations are 
suppressed by a magnetic field and are impossible.

Maxima of dependences submitted on fig. 4  are possible to identify 
with frequencies those phonons which provide a coherence condition. 
If to accept for silicon  $m^* = 0.98 m _ 0 $, where 
$ m _ 0 $ - mass at rest of electron, then the 
appropriate frequencies multiple $ 2.1\cdot10 ^ {10} $ $c^{-1}$, lay in 
limits from $ 2.1\cdot10 ^{10}~c^{-1}$ up to $ 1.26\cdot10 ^{11}~c^{-1}$ 
and correspond to acoustic fluctuations of a crystal. 
It is seen from these data  that the frequency 
$F=1.25\cdot10^{10}~c^{-1}$ accepted for 
an estimation of the minimal concentration of A- centres 
($ N _ {min} $) is less but close to $ 2.1\cdot10 ^ {10} $ $c^{-1}$.
Oscillations with the large frequencies are absent 
when B $>$ 1.3 tesla hence coherence area   can not arise 
also superconductivity is impossible. Therefore it is possible 
to expect that in magnetic fields with B $<$ 1.3 tesla the 
superconductivity caused EVC is possible but at B $>$ 1.3 tesla she
may be impossible.
\section*{4.~Experimental results and discussion}
\hspace{1.5em}  The experimental temperature dependences of 
resistance for two Si samples are submitted on fig. 5. Within the separate 
temperature areas the resistance is described by some activation 
energies which equal to energies of phonons cooperating with inherent 
oscillations of oxygen atoms. One can see on fig. 5  
when the temperature is raised up to $T =  T _ c $ the 
resistivity sharply decreases up to zero (superconductivity 
arises at $T =  T _ c $). The superconductivity exist up to rather high 
temperatures (probably up to temperature of a crystal melting and is 
higher). One can see on fig. 5 else that   curve 1 
contains two steps of resistance sharp reduction 
(at $T$ = 250K and at $T$ = $T _ c$ = 309K) but curve 2 contains one 
such step (at $T= T_c = 389.6K$). The similar characteristics 
of some samples can contain 3 or 4 such  steps. The presence 
of such steps resistance  reduction   
at approach of temperature to $T _ c$ specifies discrete (quantum) 
character of superconductivity arising. The occurrence of such steps 
can be connected with processes of arising, reorganization and 
cooperation of coherence  areas   with each other. In some samples 
alongside with the specified steps of resistance reduction the steps 
of sharp resistance increase are observed also and in such case 
superconductivity does not arise that it is possible to explain by 
deformation or destruction of  coherency areas. In any case the sharp 
changing of samples resistance at discrete temperatures can 
definitely be connected with changing of a role of various quantums 
of EVC inherent oscillations because  at absence of EVC the sharp 
jumps of resistance and superconductivity is not observed.

Becouse the given superconductivity arises due to oscillations of 
a crystal with very high frequencies appropriate to hypersound it 
is expedient to name the given superconductivity, caused by the 
electron-vibration centres, as  hyperconductivity to distinguish 
her mechanism from other mechanisms of superconductivity.

In silicon samples we observed hyperconductivity in magnetic fields 
with B up to 0.5 tesla that does not contradict an 
estimation of allowable intensity of a magnetic field.
Hyperconductivity existed at density of a current up to 
10 $ A/cm ^ 2 $  that not contradicts estimations of critical 
density for a current density.

The occurrence hyperconductivity is connected to occurrence   
of thermopower (TEP) features. On fig.6 are given  temperature 
dependences of TEP for the  sample designated by 2 on fig. 5. 
Such alternating-sign of temperature dependence TEP is 
characteristic for  normal-hyperconducting transition. 
It seems  the behaviour TEP reflects complex dynamic processes 
of formation and interaction of coherence areas  among themselves.

Occurrence of superconductivity is accompanied by huge increase  
of  thermal conductivity. In accordance with  experimental 
data the meaning of thermal conductivity in some samples  increase 
more than $ 10 ^ 5 $ times. Probably the hyperconductivity is 
accompanied by heat superconductivity.  

Hyperconductivity arises in various crystals containing EVC. 
On fig. 7 the available experimental data about meanings 
of the $T_c$ in various hyperconductor samples are submitted. 
One can see on  fig. 7 that normal-hyperconducting transition 
temperature $T_c$ shows the tendency to reduction at increasing of 
average size of the basic substance nuclear number ($Z_{mid}$).
Straight linees a and b on fig. 7 restrict the area in which the 
experimental points are placed. One can see on fig. 7 that the 
experimental meanings $ T _ c $ for every material can change 
over a wide range between  linees a and b. On the other hand 
situation of straight linees a and b differs on insignificant 
size of energy $kT\simeq17$~meV appropriate to acoustic phonons. 
It speaks about importance not only inherent oscillations of EVC 
but  acoustic phonons simultaneously for formation of 
hyperconductivity in different materials.
The minimal experimental meanings of $ T _ c $  for Si and Ge are 
in agreement  with given estimation of $T_ c$ when $\mu$=2 for 
$\alpha$-inherent oxygen oscillator. According to Eq.~(\ref{7}) 
the increase S corresponds to reduction $T_c$ and  promotes 
hyperconductivity existence in wide area of temperatures.  
The meanings $ S\gg1 $ are necessary for existence hyperconductivity from
near to room temperatures up to higher temperatures that it is possible 
to provide by introducing EVC in crystals. The experimental data about 
meanings $T_c$ for semiconductors with small  $E_g$ can be coordinated 
with Eq.~({\ref{7}) only by taking into account $\beta$-type or $\gamma$-type 
of inherent oscillations because solutions of Eq.~({\ref{7}) with 
participation of $\alpha$ - inherent oscillations are absent.
 
\section*{5.~Conclusion}
The superconductivity caused by electron-vibrational 
centres - hyperconductivity can exist in the field of room 
temperatures and at higher temperatures in various materials in wide 
electrical current density and magnetic intensity intervals. 
That allows ones to hope on opportunity   
of hyperconductivity  application in science and engineering.
\vspace{0.5cm}

\newpage
\section *{Figure captions}
\newcommand{\Fig}[2]{\noindent{\bf Fig.~#1.~}
\parbox[t]{15cm}{\baselineskip 24pt #2}\\[5mm]}
\Fig{1}
{Experimental spectrum IR reflection of polarized light from a surface 
of quartz (1).  The calculated spectrum of reflection by "zero" 
inherent  fluctuations of oxygen (2). }
\Fig{2}
{Spectrum of photoconductivity (1) and spectrum of IR  transmittance (2) 
áaused by photoionization of A - centres in silicon at T = 80K. 
The arrows designate the appropriate each other minima 
of spectra and the number participating phonons is specified by figures.
On an insertion the experimental data about splitting LO and TA 
phonons are given depending on concentration of A - centres.}
\Fig{3}
{'haracteristics volt - capacitance  of contact Si - Al containing 
A - centres are measured at room temperature $ T < T _ c $ on various frequencies: 
0.2 MHz~~(1); 0.5 MHz~~(2); 1.0 MHz~~(3); 5.0 MHz~~(4); 10.0 MHz~~(5); 
20.0 MHz~~(6).}
\Fig{4}
{Derivative of  voltage on Si  sample with respect to  
magnetic intensity (B) at room temperature $ T < T _ c $.}
\Fig{5}
{Temperature dependences of resistance for two of Si  samples 
containing  various concentration of A - centres: 
$N_1~\approx~5\cdot10^{14}~cm^{-3}$ and $N_2~\approx~6\cdot10^{15}~cm^{-3}$.}
\Fig{6}
{Temperature dependence of thermopower for the same Si sample whose 
characteristic is designated on fig. 5  by number 2.}
\Fig{7}
{The experimental meanings of normal-hyprconducting transition 
temperatures ($T_c$) according to average meanings of nuclear
number ($Z _ {mid}$) in various samples containing electron-vibrational 
centres. Straight linees a and b limit area in which the experimental 
points lay.}

\begin{thebibliography}{99}
\bibitem{Onn11} 
              Onnes H.K., Leiden. Comm., 1911, p.~120b,  122b.        
\bibitem{Gin77}
              V.L.Ginzburg, E.G.Maksimov. M. Science, 1977, in russian. 
\bibitem{Gin92}
              V.L.Ginzburg, E.G.Maksimov. Superconductivity. Physics, 
              chemistry, engineering. 1992, v.~5,  ü9, p.~1543, in russian.         
\bibitem{Bar57}
              J.Bardeen, L.N.Cooper, J.R.Schrieffer. Phys. Rev., 
              1957, v.~108, p.~1175.        
\bibitem{Vdo96}
              V.A.Vdovenkov. Proceedings of MIREA, 1996, p.~148, in russian.         
\bibitem{Vdo99}
              http://xxx.lanl.gov/cond-mat/9904299.         
\bibitem{Born27}
               M.Born, R.Oppenheimer. Ann.~D.~Phys. 1927,  v.~84, N4, p.~457.         
\bibitem{Dav73}
              A.S.Davidov. The quantum mechanics. M. Science, 
              1973, p.~613, in russian.         
\bibitem{Lax55}
              M.Lax, E Burstein. Phys.~Rev. , 1955, v.~100,  p.~592.         
\bibitem{Hard62}
               J.R.Hardy, S.D.Smith, W.Taylor. Proc. Int.conf. 
               on Phys. of Semicond., Exter, 1962, p.~521.         
\bibitem{Hua50}
              K.Huang and A.Rhys. Proc.~Roy.~Soc., 1950, v.~A204, p.~406.         
\bibitem{Pek51}
              S.I.Pekar. Researches under the electronic theory of crystals. 
              Gostechizdat, 1951, in russian.         
\bibitem{Pek52}
              S.I.Pekar. JETF, v.~22, 6, 1952, p.~641, in russian.         
\bibitem{Pek53}
              S.I.Pekar. UFN, v.~L, 2, 1953, p.~197, in russian.        
\bibitem{ORo53}
              R.O'Rourke. Phys.~Rev., 1953, v.~91, p.~265.
\bibitem{Kai56}
              W.Kaiser, P.H.Keck, G.F.Lange. Phys.~Rev., 1956,  
              v.~101, N4, p.~1264.
\bibitem{Kai57}
              W.Kaiser, P.H.Keck.. Jorn.~Appl.~Phys., 1957, v.~28, 
              N8, p.~882.
\bibitem{Hro57}
              H.J.Hrostowski, R.H.Kiser. Phys.~Rev., 1957, v.~107, 
              N4, p.~966.
\bibitem{Abe66}
              T.Abe, S.Maruyama. Jap.~J.~Appl.~Phys., 1966, v.~5,  
              N10,  p.~979.
\bibitem{Vla70}
              Infra-red spectra of alkaline oxides. Editors: A.G.Vlasov  
              and V.A.Florinskaya, St.Petersburg, Chemistry, 
              1970, in russian.
\bibitem{Vdo81}
              V.A.Vdovenkov, N.V.Korotkova. Electronic engineering, 
              ser.~6 "materials", 1981,  rel.~6(155), p.~54, in russin.
\bibitem{Hua51}
              K.Huang. Proc.~Roy.~Soc., 1951, A208, p.~352.

\bibitem{Born54}
              Born M., Huang K. Dynamical Theory of Crystal Lattices, 
              London-New York, 1954, p.~82.
 
\bibitem{Ros51}
              Rosenfeld L. Theory of Electrons, Norh Holland, 
              Amsterdam, 1951.
\bibitem{Nor58}
              Norieres P., Pines D. Phys.~Rev., 1958, v.~109, p.~741.
\bibitem{Bem59}
              G.Bemski. J. Appl.~Phys., 1959, v.~30, p.~1195.
\bibitem{Aki68}
              I.P.Akimchenko, V.S.Vavilov, V.A.Vdovenkov, A.F.Plotnilov. 
              Fiz.~Tverd.~Tela, 1968, v.~10, p.~3677, in russian.
\bibitem{Gor63}
              I.Goroff and L.KIeinman. Phys.~Rev. , 1963, v.~132, 
              N3, p.~1080. 
\bibitem{Mor57}
              F.J.Morin, T.H.Geballe and Herring. Phys.~Rev., 
              1957,  v.~101, p.~525.
\bibitem{Fre53}
              H.P.R.Frederikse. Phys.Rev., 1953, v.~91, p.~491;~ 
              1953, v.~92, p.~248.
\bibitem{Herr58}
               C.Herring. Halbleiter und Phosphore 
               (eds. M.~Schon, H.~Welker), Vieweg, Braunschweig, 
               1958, p.~184.         
\end{thebibliography}
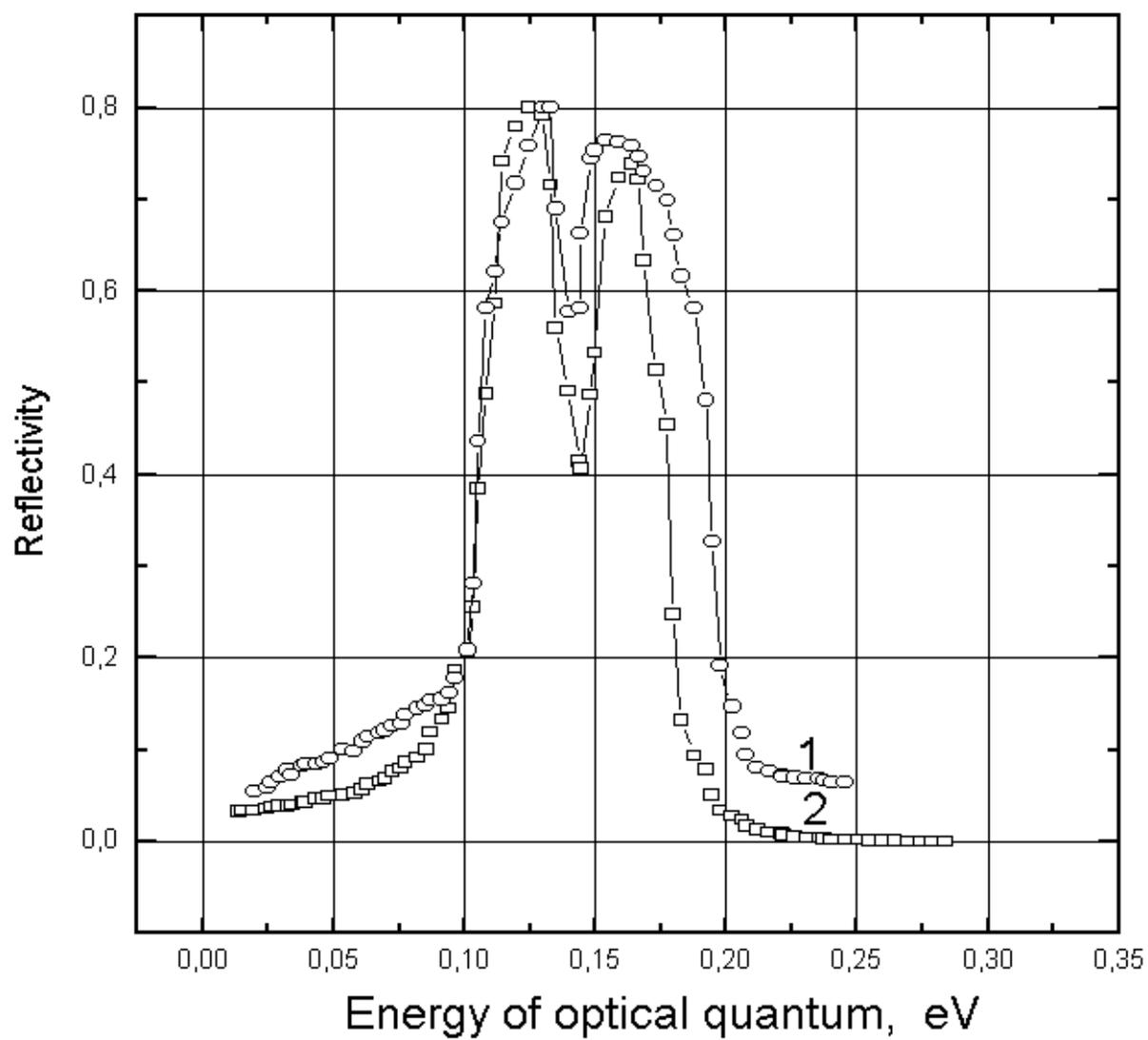
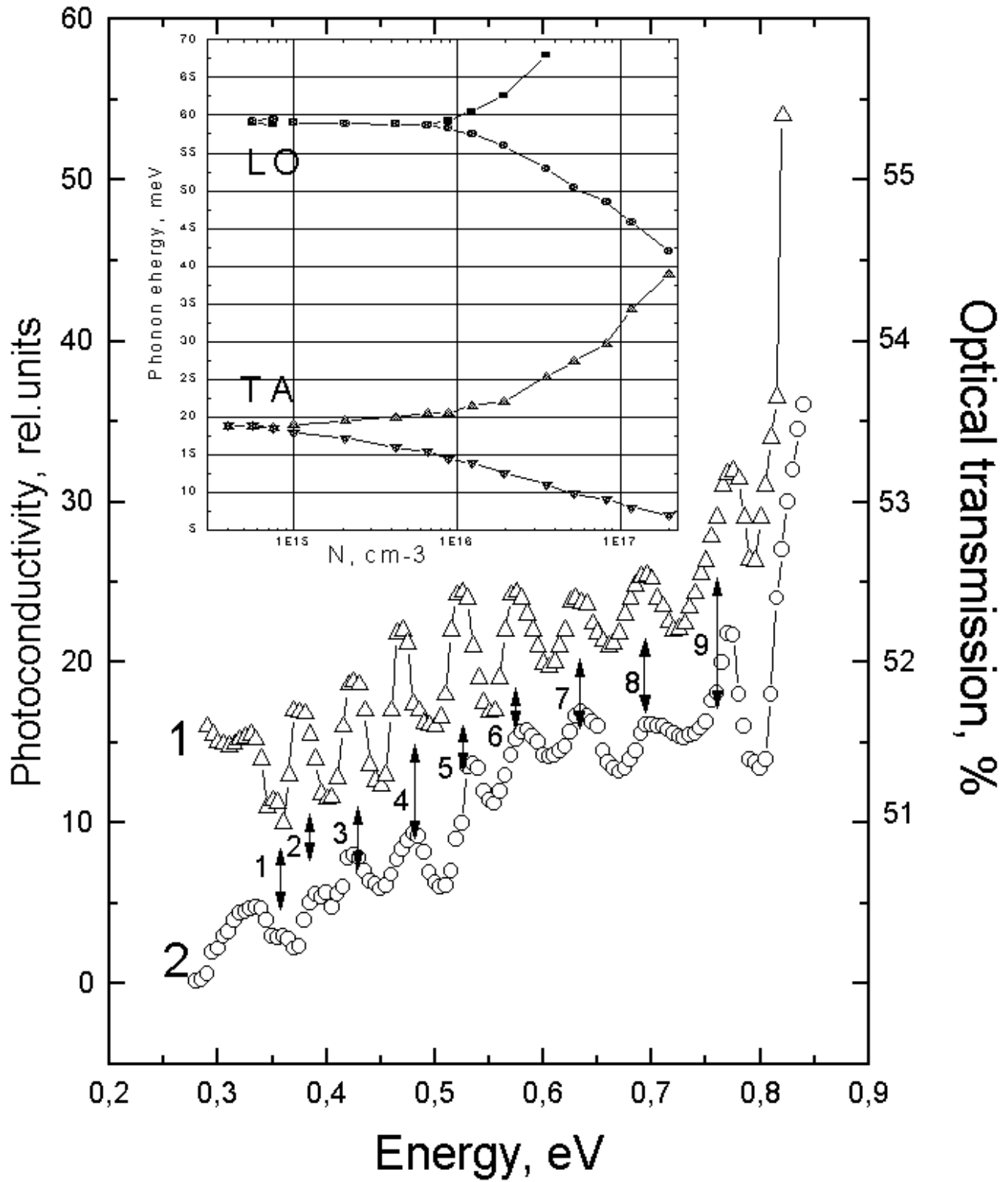
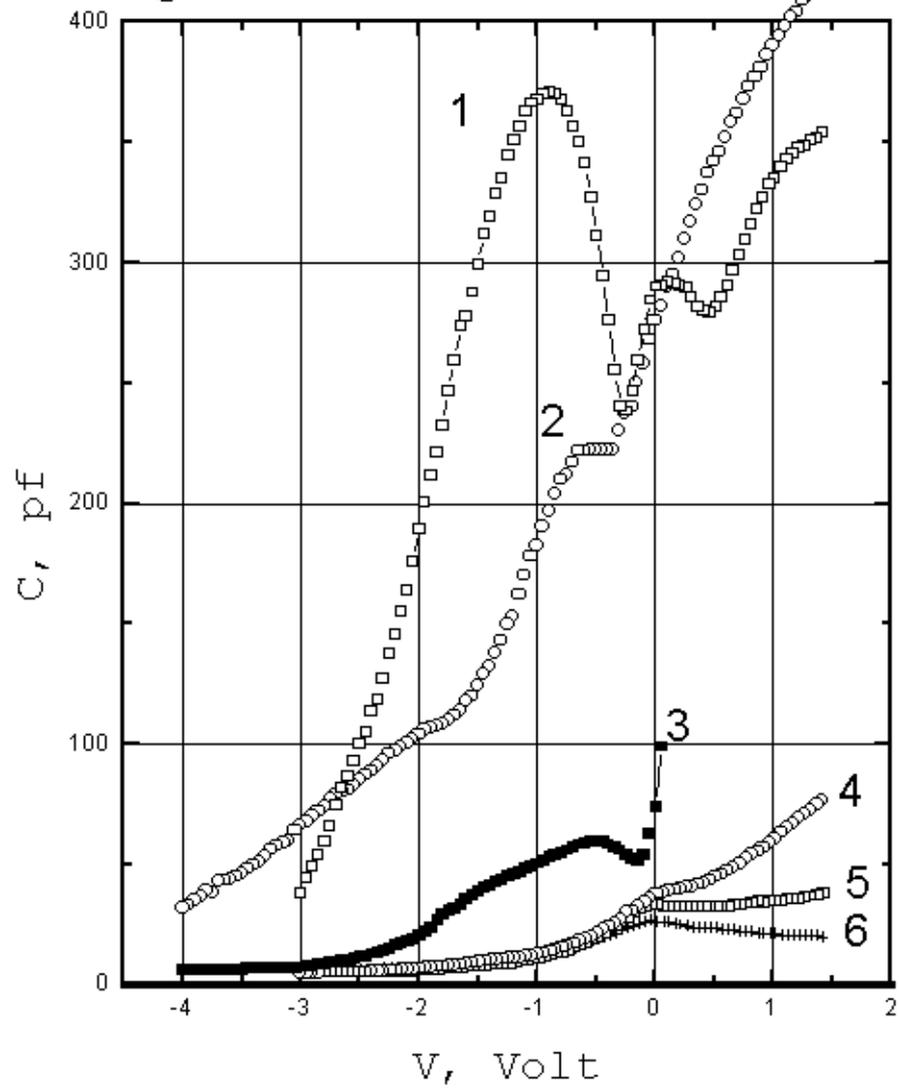
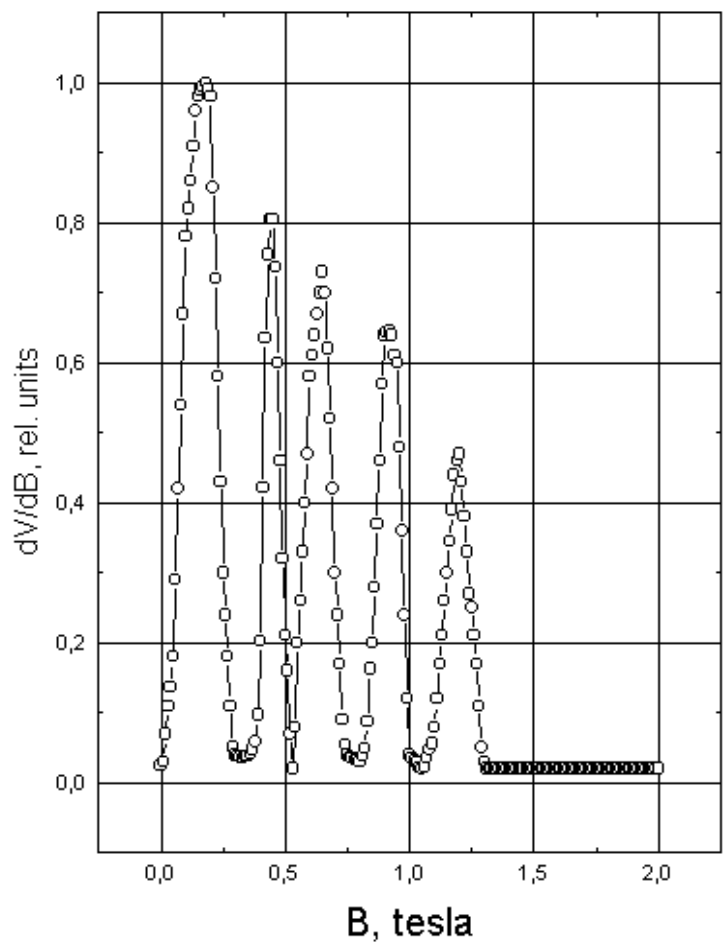
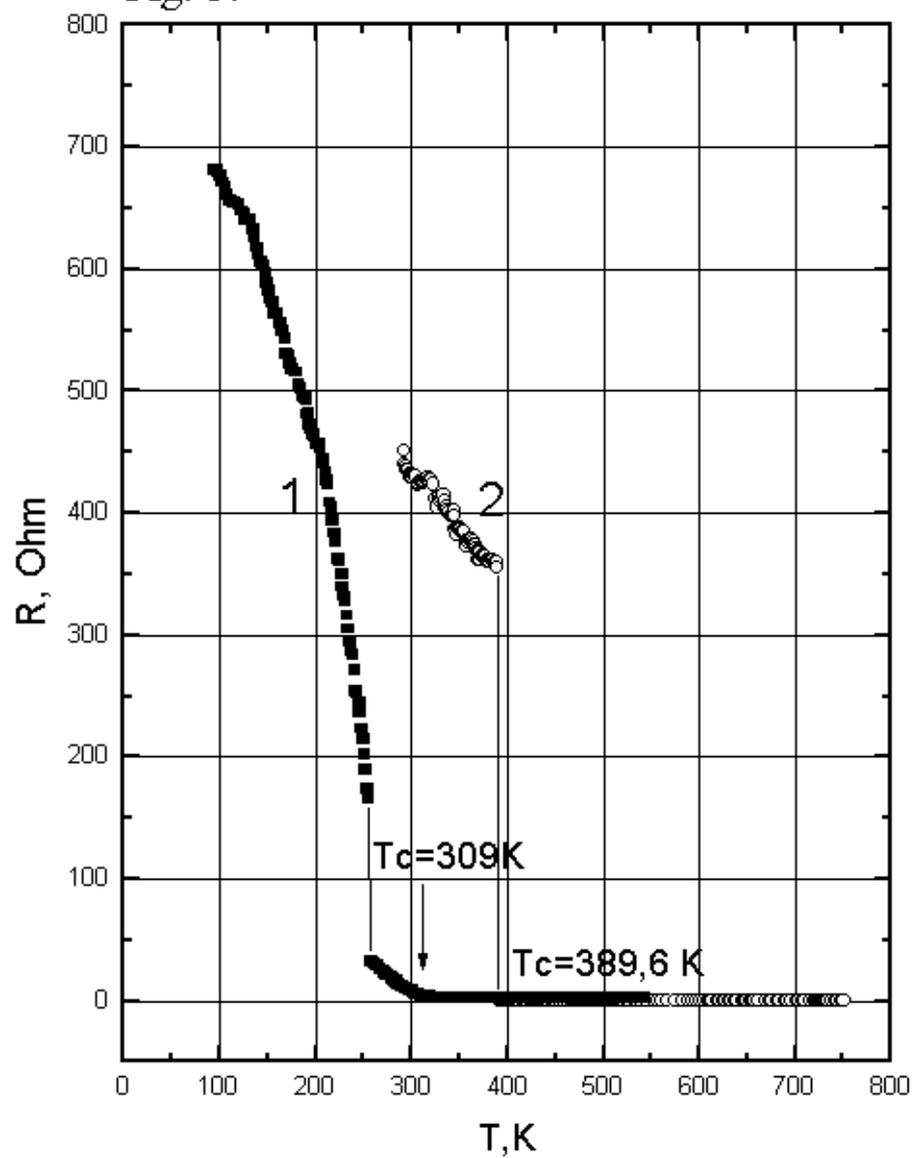
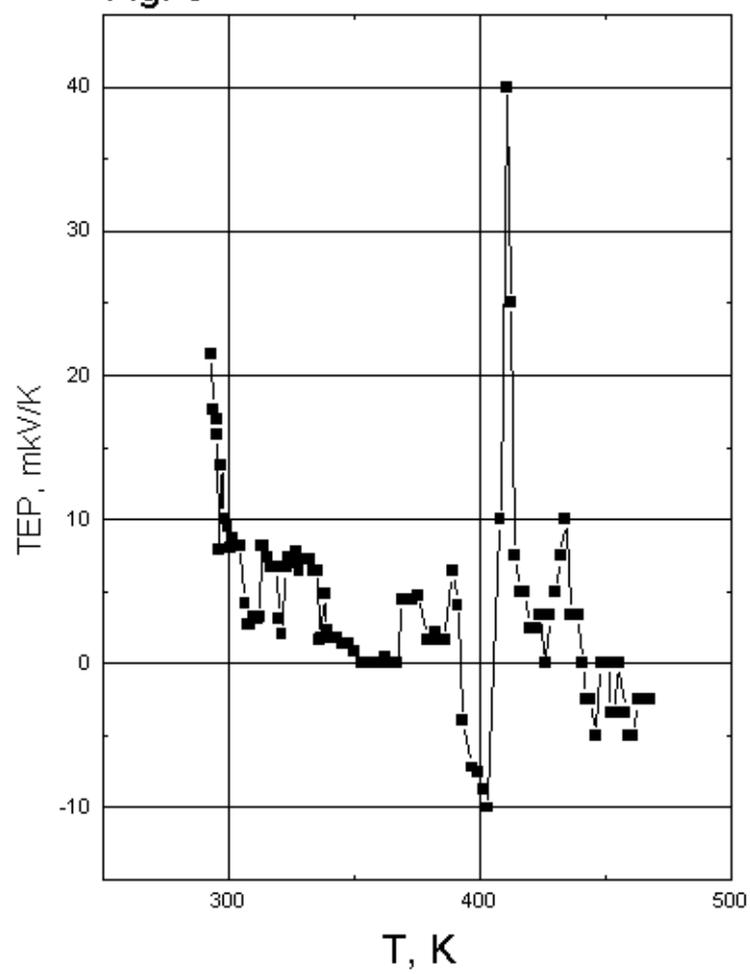
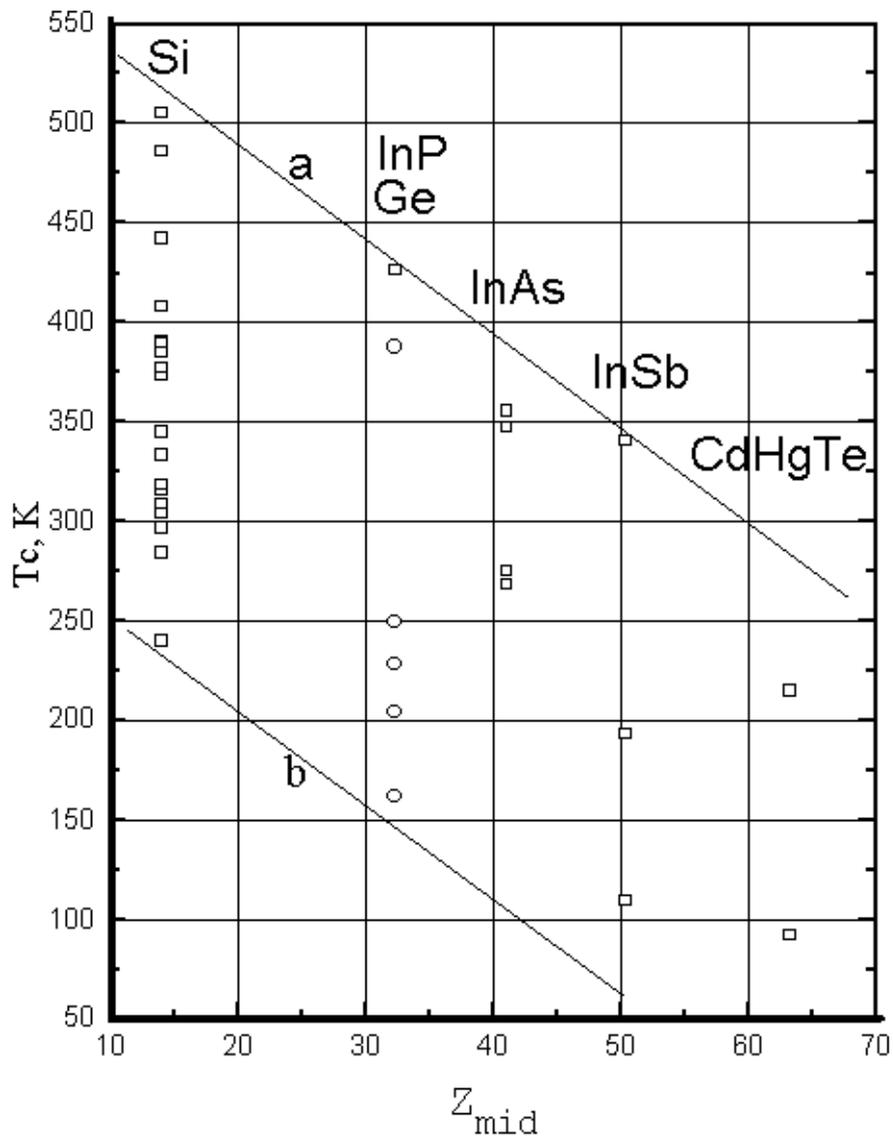
\end{document}